\documentclass{appolb}
\usepackage{graphicx}
\usepackage{amsmath}

\begin{document}
\title{Higher order cumulants of electric charge and strangeness fluctuations on the crossover line.%
\thanks{Presented at Criticality in QCD and the Hadron Resonance Gas, 29-31 July 2020, Wroclaw Poland}%
}
\headtitle{Electric charge fluctuations ~~ printed on \today}
\headauthor{D. Bollweg ~~ printed on \today}
\author{D. Bollweg\thanks{speaker},  J. Goswami, F. Karsch, C. Schmidt
\address{Fakult\"at f\"ur Physik, Universit\"at Bielefeld, D-33615 Bielefeld, Germany}\vspace{5mm}
\\
{S. Mukherjee
}
\address{Physics Department, Brookhaven National Laboratory, Upton, New York 11973, USA}
}
\maketitle
\begin{abstract}
We present lattice QCD calculations of higher order cumulants of electric charge distributions for small baryon chemical potentials $\mu_B$ by using up to NNNLO Taylor expansions. Ratios of these cumulants are evaluated on the pseudo-critical line, $T_{pc}(\mu_B)$, of the chiral transition and compared to corresponding measurements in heavy ion collision experiments by the STAR and PHENIX Collaborations. We demonstrate that these comparisons give strong constraints on freeze-out parameters. Furthermore, we use strangeness fluctuation observables to compute the ratio $\mu_S/\mu_B$ on the crossover line and compare it to $\mu_S/\mu_B$ at freeze-out stemming from fits to strange baryon yields measured by the STAR Collaboration.
\end{abstract}
\PACS{11.15.Ha, 12.38.Gc, 12.38.Mh, 24.60.-k}
  
\section{Introduction}
Uncovering the phase structure of QCD poses a long-standing, open challenge in heavy ion research. Large efforts - in the form of relativistic heavy ion collision experiments - are made to find signs of a critical point proposed by many model calculations. This point would provide an important landmark in the largely unknown phase diagram of QCD and signals itself for instance through divergences in conserved charge fluctuations that couple to the order parameter. Remnants of this divergence might be seen in measurements of higher order cumulants of conserved charge distributions in heavy ion collisions if freeze-out occurs in the vicinity of the critical point. Baryon number and strangeness fluctuations are studied in heavy ion collision experiments through the measurement of Event-by-Event fluctuations of proxy particle species, such as proton or kaon numbers, respectively. Electric charge fluctuations on the other hand can be measured without resorting to proxies, making them particularly attractive for comparisons with lattice QCD calculations. Here, we want to provide thermal QCD baselines for higher order cumulants of electric charge fluctuations via state-of-the-art lattice QCD calculations and contrast them with results from the STAR and PHENIX experiments. Furthermore, we will also use strangeness fluctuation observables to construct $\mu_S/\mu_B$ on the crossover line. 

\section{Setup}
In previous studies \cite{bazavov2020skewness}, we presented calculations of higher order cumulants of net-baryon number fluctuations based on a high statistics data set of (2+1)-flavor HISQ gauge field configurations with physical light and strange quark masses. We use degenerate light quark masses and a light to strange quark mass ratio $m_l/m_s=1/27$. Here we add to this analysis the corresponding higher order cumulants of electric charge and strangeness fluctuations.
By computing up to eighth order generalized susceptibilities
\begin{align}
  \label{eq:chilog}
\chi^{BQS}_{ijk}(T,\vec{\mu})=\frac{1}{VT^3}\frac{\partial^{i+j+k}\ln{Z(T,\vec{\mu})}}{\partial\hat{\mu}_B^i\partial\hat{\mu}_Q^j\partial\hat{\mu}_S^k},\;\;\mathrm{with} \;\hat{\mu}_X=\frac{\mu_X}{T}
\end{align}
we are able to construct Taylor series coefficients $\tilde{\chi}^{X,k}_{n}(T)$ for $n$-th order strangeness and electric charge cumulants
\begin{align}
    \chi^{X}_{n}(T,\mu_B)=\sum_{k=0}^{k_{\mathrm{max}}}\tilde{\chi}_{n}^{X,k}(T)\hat{\mu}_{B}^{k}, \;\;\mathrm{with}\;\; X=Q,S.
\end{align}
In order to match the conditions present in heavy ion collisions, we constrain the Taylor series such that the ratio of charge density $n_Q$ to baryon density $n_B$ is $n_Q/n_B=0.4$ and the strangeness density is $n_S=0$. This is achieved by expanding electric charge and strangeness chemical potentials in $\mu_B$ with coefficients $q_i$ and $s_i$ chosen such that $n_Q/n_B=0.4, \;n_s=0$ hold at each order,
\begin{align}
  \label{eq:museries}
  \hat{\mu}_Q(T,\mu_B)&=\sum_{i}q_{2i+1}(T)\hat{\mu}_B^{2i+1},\\\nonumber
  \hat{\mu}_S(T,\mu_B)&=\sum_{i}s_{2i+1}(T)\hat{\mu}_B^{2i+1}.                        
\end{align}
Explicit formulas for $q_j$ and $s_j$ for $j \leq 5$ can be found in Appendix B of \cite{eospaper}.
A full list of the expressions for $\tilde{\chi}_{n}^{X,k}$ will be given in an upcoming publication.
Finally, we form cumulant ratios
\begin{align}
  \label{eq:rdef}
  R^{X}_{nm}(T,\mu_B)=\frac{\chi^{X}_{n}(T,\mu_B)}{\chi^{X}_{m}(T,\mu_B)}=\frac{\sum_{k=0}^{k_{\mathrm{max}}}\tilde{\chi}_{n}^{X,k}(T)\hat{\mu}_{B}^{k}}{\sum_{l=0}^{l_{\mathrm{max}}}\tilde{\chi}_{m}^{X,l}(T)\hat{\mu}_{B}^{l}},
\end{align}
in order to cancel the volume factor in \eqref{eq:chilog} that is unknown in heavy ion collision experiments. In this work, we will focus on the mean-to-variance ratio $R^{X}_{12}=\frac{M_X}{\sigma_X}$, the skewness ratio $R^{X}_{31}=\frac{S_X\sigma^3_X}{M_X}$ and the kurtosis ratio $R^X_{42}=\kappa_X\sigma_X^2$. We evaluate \eqref{eq:rdef} for nine temperatures ranging from 135 MeV to 175 MeV and normalized chemical potentials $\hat{\mu}_B$ ranging from 0 to 2 in steps of 0.01.
\begin{figure}[htb]
\centerline{%
\includegraphics[width=12.5cm]{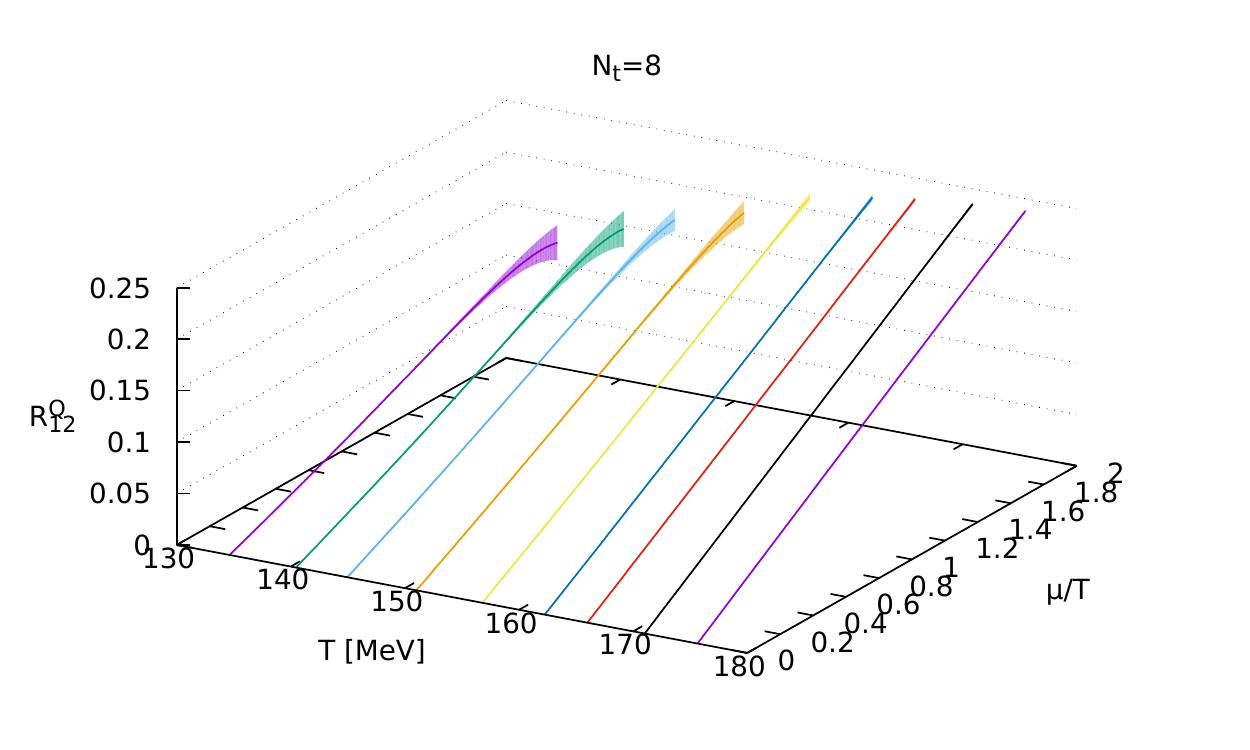}}
\caption{$R^Q_{12}$ for $N_t=8$ in the $(T,\hat{\mu}_B)$-plane.}
\label{fig:muscan}
\end{figure}
This produces nine slices in the $(T,\hat{\mu}_B)$-plane that trace out the $R^X_{nm}(T,\hat{\mu}_B)$ surface. An example for this is shown in Fig. \ref{fig:muscan} for the case of $R^{Q}_{12}$ for $N_t=8$ lattices. The data for the different lattice sizes is then jointly fitted via low-order polynomial Ans\"atze to obtain continuum extrapolations. The details of the fitting procedure will be described in a forthcoming publication.
\section{Electric charge fluctuations on the crossover line}
\begin{figure}[htb]
\centerline{%
\includegraphics[width=12.5cm]{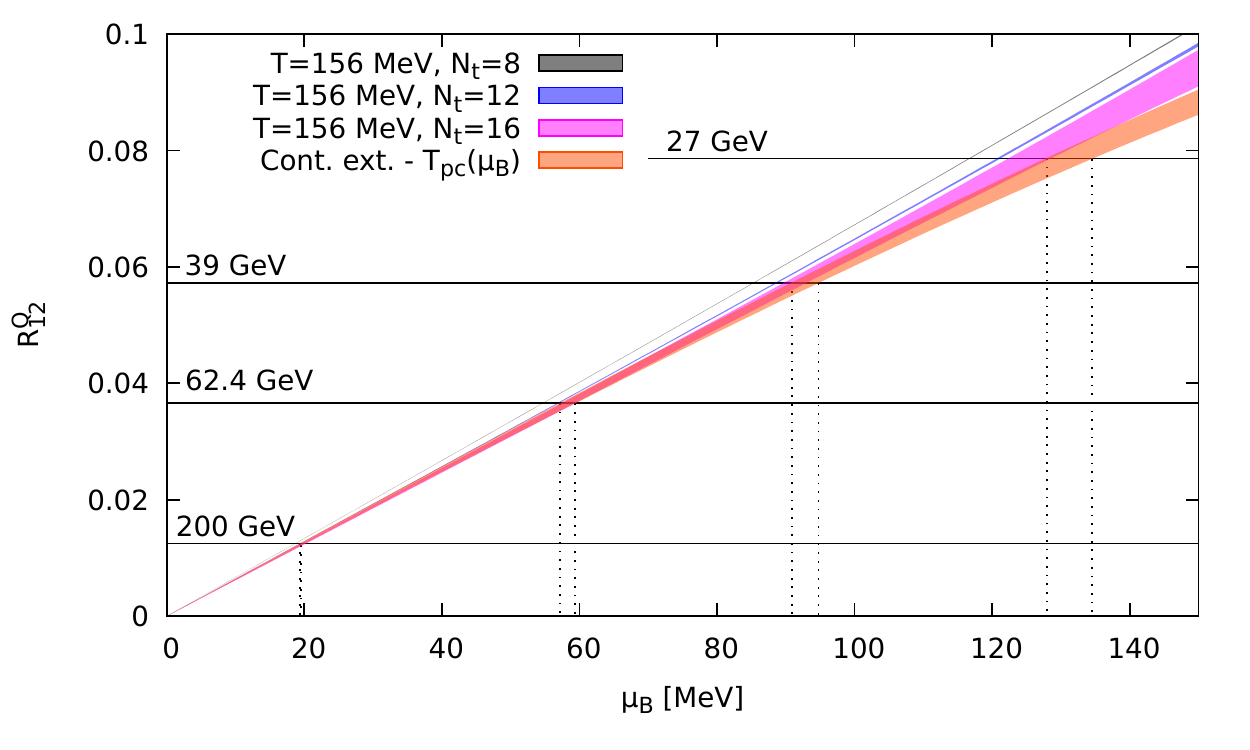}}
\caption{$R^{Q}_{12}$ for different $N_t$ and continuum extrapolation along $T_{pc}(\mu_B)$. Horizontal lines show results from STAR \cite{StarNetChargeDist} at different beam energies.}
\label{fig:rq12star}
\end{figure}

The mean to variance ratio $R^{Q}_{12}=\frac{M_Q}{\sigma^2_Q}$, shown across the $(T,\hat{\mu}_B)$-plane in Fig. \ref{fig:muscan}, has been calculated to NNNLO in $\mu_B$. Apart from a small area in the low $T$ and high $\hat{\mu}_B$ region, it shows nearly perfect linear behavior in $\hat{\mu}_B$ direction with very small variation along the $T$ direction. This qualifies it to be used as a baryometer when comparing this quantity with experimental results.
In Fig. \ref{fig:rq12star}, we show $R^{Q}_{12}$ for the three different lattices sizes at temperatures close to $T_{pc,0}$ as well as its continuum extrapolation evaluated on the crossover line
    \begin{align*}
T_{pc}(\mu_B)=T_{pc,0}\left(1-\kappa^{B,f}_{2}\left(\frac{\mu_B}{T_{pc,0}}\right)^2\right)
    \end{align*}
with $T_{pc,0}=156.5\pm 1.5\; \mathrm{MeV}$ and $\kappa^{B,f}_{2}=0.012 \pm 0.004$ \cite{Bazavov_2019}.
This figure also contains the mean to variance ratios measured by STAR at different beam energies depicted as horizontal lines. The intersections of the experimentally determined values with the lattice result on the crossover line are highlighted with dashed vertical lines. These marks provide a mapping between beam energies and freeze-out chemical potentials if that freeze-out happens on the crossover line. The numerical values extracted from this comparison are listed in Table \ref{tab:muf}.
\begin{table}[htb]
  \centering
  \begin{tabular}{c|c}
    $\sqrt{s}_{NN}$ [GeV] & $\mu_{B,f}$ [MeV]\\\hline
    200 & 19.4(1) \\
    62.4 & 58(1) \\
    39 & 92(2) \\
    27 & 131(3)
  \end{tabular}
  \caption{Chemical potentials extracted from comparing results from STAR \cite{StarNetChargeDist} with lattice QCD.}
  \label{tab:muf}
\end{table}

\begin{figure}[htb]
\centerline{%
  \includegraphics[width=6.5cm]{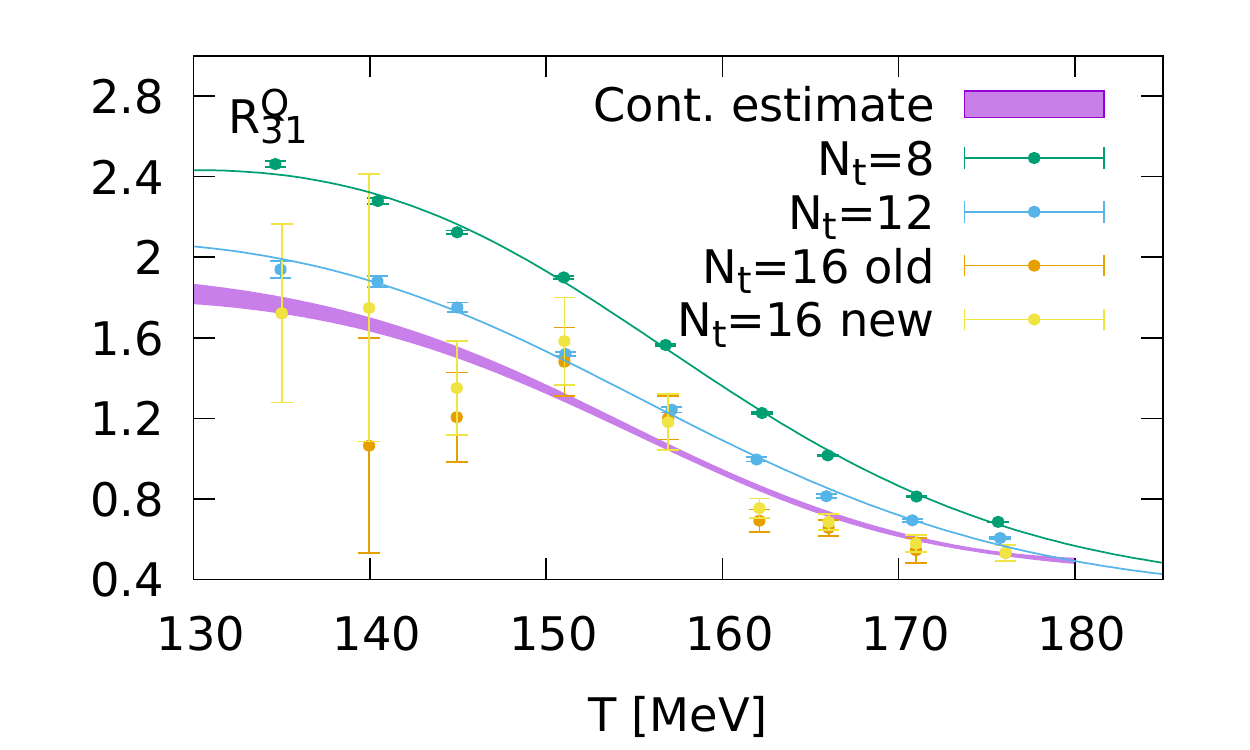}
\includegraphics[width=6.5cm]{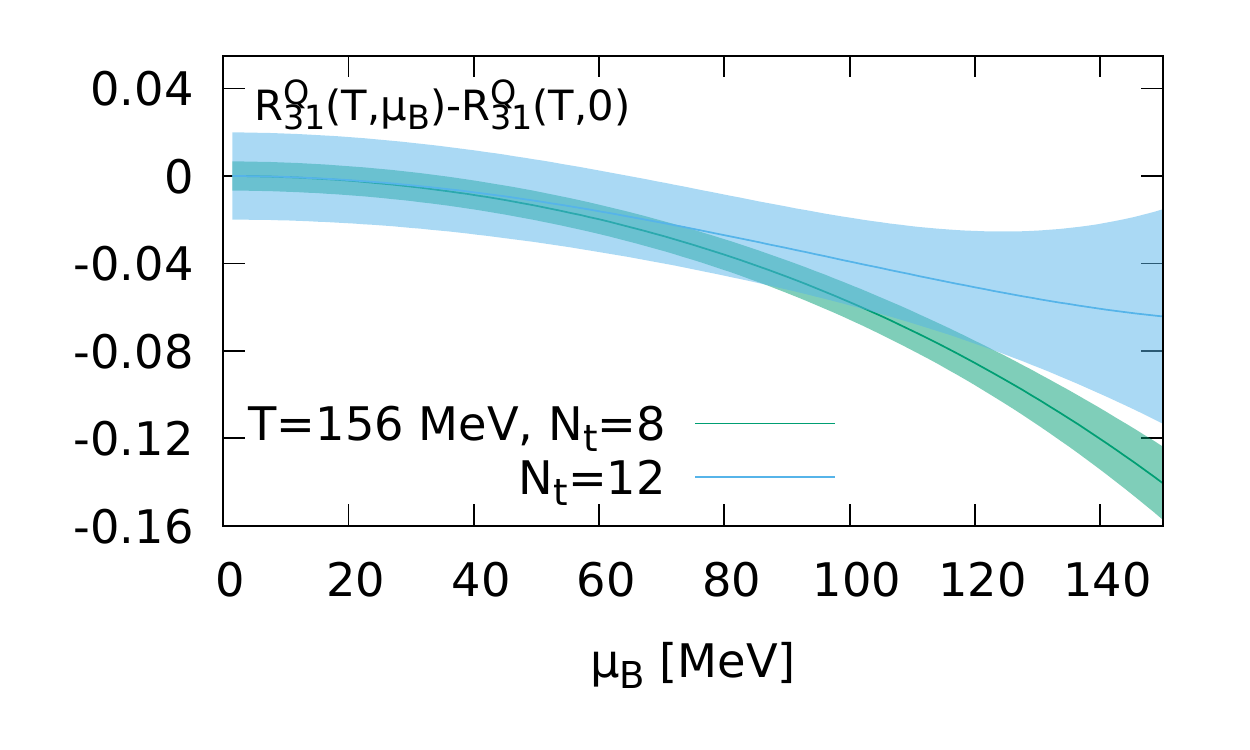}}
\caption{{\it Left:} $R^Q_{31}(T,\mu_B=0)$ for different lattices sizes. {\it Right:} $\mu_B$ dependence of $R^Q_{31}$ for $N_t=8,12$ lattices.}
\label{fig:rq31}
\end{figure}

\begin{figure}[htb]
  \centerline{%
    \includegraphics[width=12.5cm]{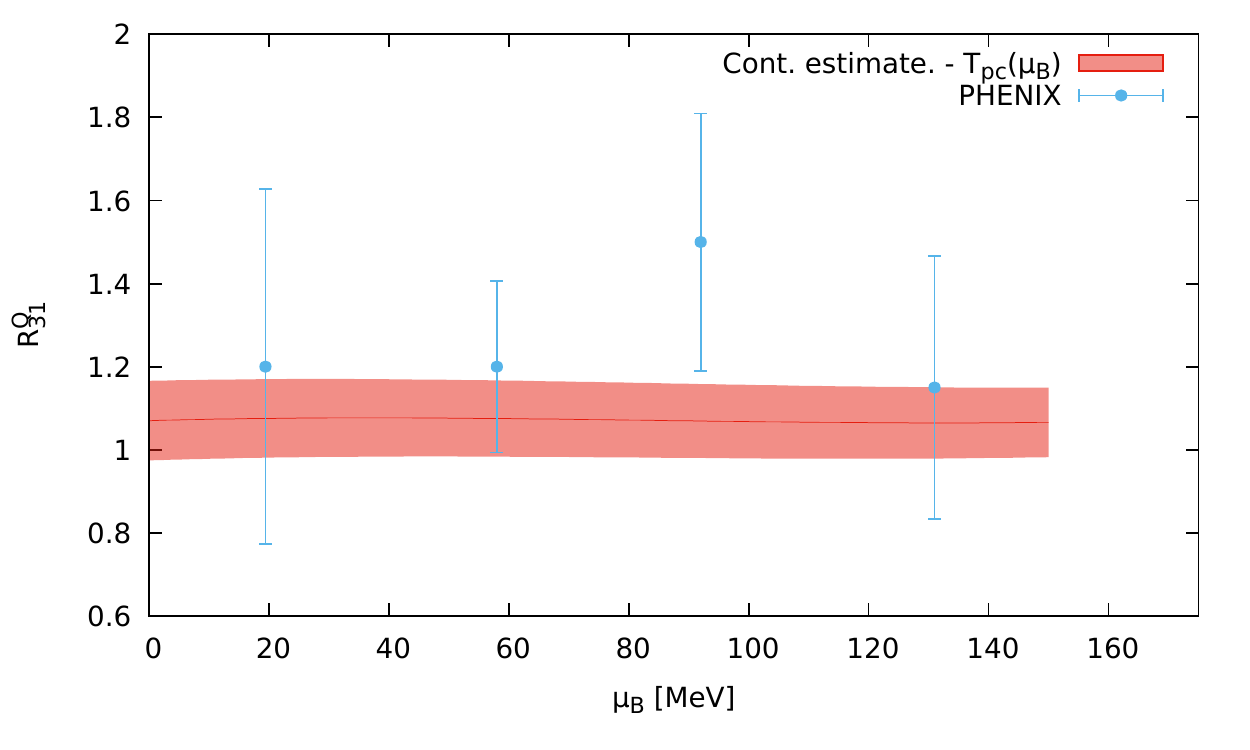}}
\caption{Comparison of the continuum estimate of $R^Q_{31}(T_{pc}(\mu_B),\mu_B)$ with results from PHENIX \cite{phenix}.}
\label{fig:rq31mu}
\end{figure}

\begin{figure}[htb]
\centerline{%
  \includegraphics[width=6.5cm]{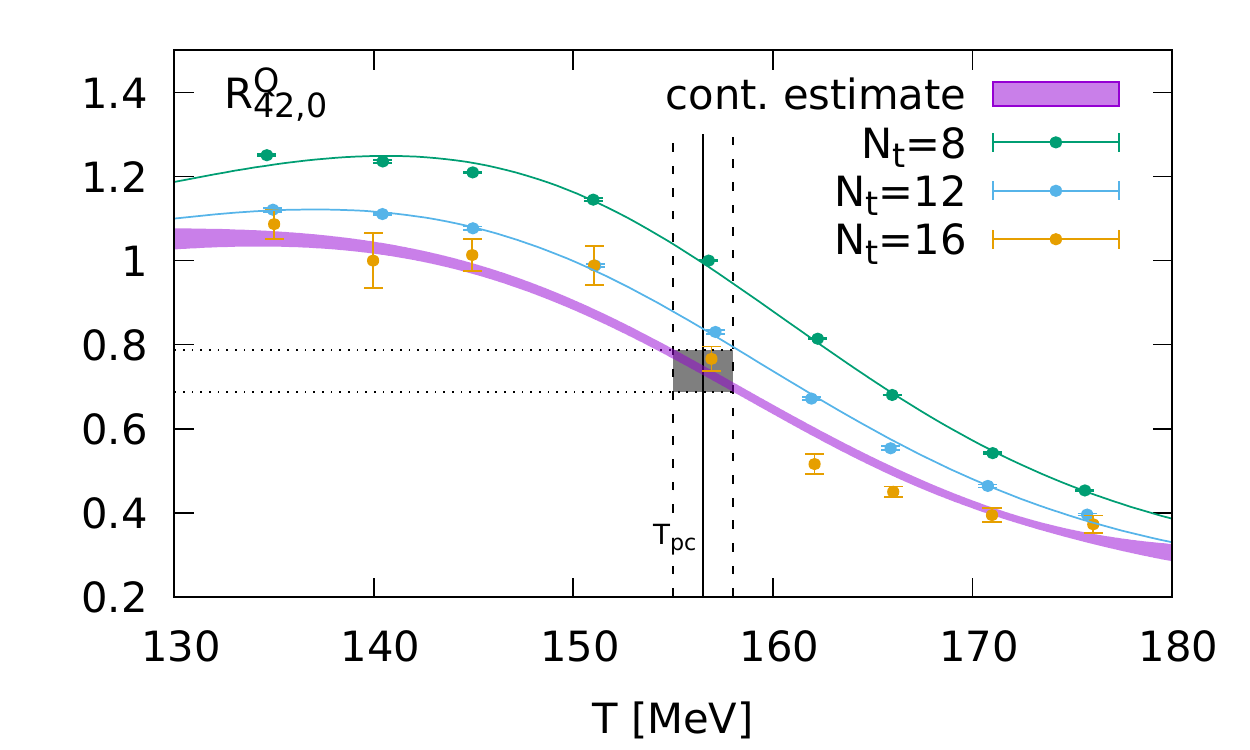}
\includegraphics[width=6.5cm]{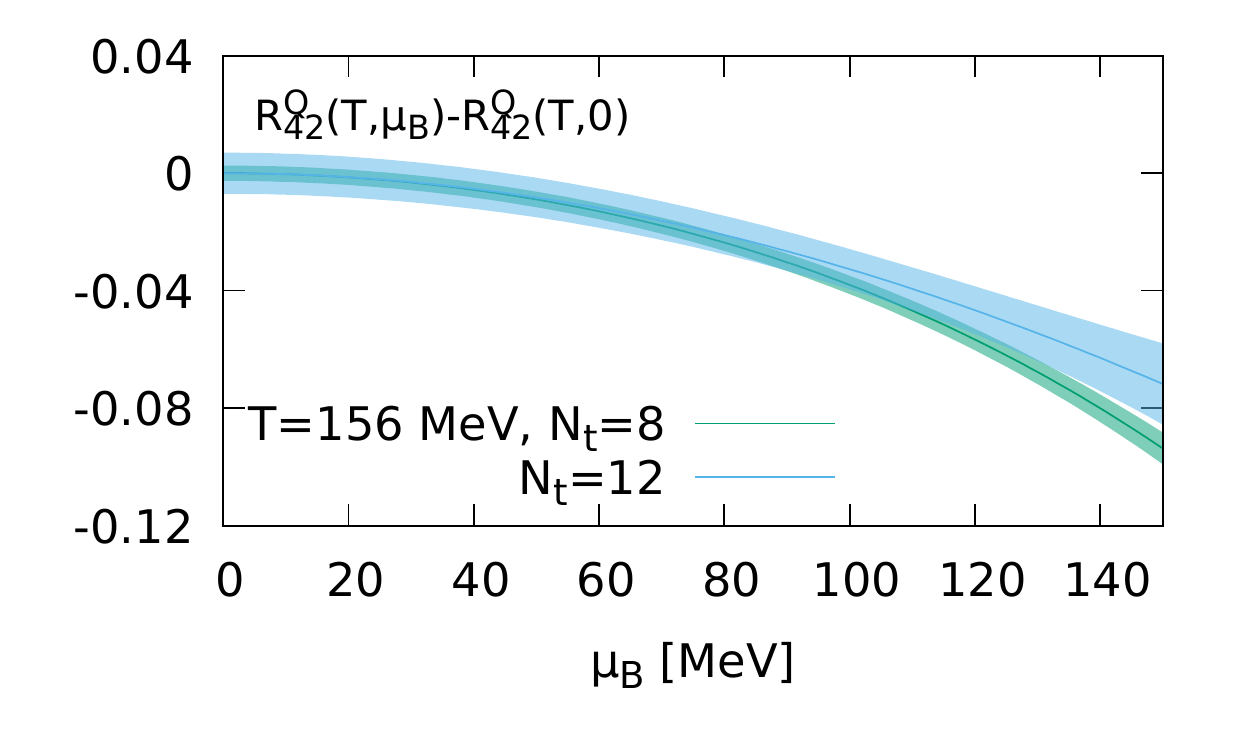}}
\caption{{\it Left:} $R^Q_{42}(T,\mu_B=0)$ for different lattices sizes. {\it Right:} $\mu_B$ dependence of $R^Q_{42}$ for $N_t=8,12$ lattices.}
\label{fig:rq42}
\end{figure}

We also calculated the skewness ratio $R^{Q}_{31}(T,\mu_B)=\frac{S_Q\sigma^3_Q}{M_Q}$ to NNLO in $\mu_B$. As seen in Fig. \ref{fig:rq31}, the roles of $T$ and $\mu_B$ are reversed when compared to $R^Q_{12}$. The skewness ratio shows a strong variation with temperature but only a mild dependence on chemical potential. In $T$ direction, $R^Q_{31}$ possesses a (mirrored) sigmoidal shape and decreases with increasing temperature while in $\mu_B$ direction, $R^Q_{31}$ decreases only mildly with increasing $\mu_B $. $R^Q_{31}$ therefore functions as a thermometer when comparing lattice results with measurements from heavy ion collisions. By evaluating the continuum estimate on the crossover line, the two effects cancel such that $R^Q_{31}(T_{pc}(\mu_B),\mu_B)$ remains almost constant when varying $\mu_B$. We find $R^Q_{31}(T_{pc}(\mu_B),\mu_B)=1.07(9)$ for $\mu_B<150$ MeV. 
 The current continuum estimate of $R^Q_{31}$, based on our $N_t=8$ and $N_t=12$ data, evaluated on the crossover line is shown as a red curve in Fig. \ref{fig:rq31mu} together with the skewness ratio measured by the PHENIX experiment \cite{phenix} in a pseudorapidity range of $\vert\eta\vert \leq 0.35$ and $27\;\mathrm{MeV} \leq \sqrt{s_{NN}} \leq 200 \;\mathrm{MeV}$ depicted as blue data points. Here we used the $\mu_B$-values determined by comparing to the STAR results and given in Table \ref{tab:muf} to plot the PHENIX data for $R^Q_{12}$.  Fig. \ref{fig:rq31mu} shows that the PHENIX results are consistent with lattice QCD calculations on the crossover line.
The kurtosis ratio $R^Q_{42}=\kappa_Q\sigma_Q^2$ has also been calculated to NNLO in $\mu_B$. Its behavior in the $(T,\mu_B)$-plane is very similar to $R^Q_{31}$, albeit with a different magnitude. The results are shown in Fig. \ref{fig:rq42}.  The error of $R^Q_{42}$ is noticeably smaller than that of the skewness ratio since it does not contain the noisy Baryon-electric charge correlations that plague $R^Q_{31}$. We estimate $R^Q_{42}(T_{pc})=0.73(5)$ for $\mu_B<150$ MeV. Unfortunately, measurements of the kurtosis ratio in heavy ion collision experiments come with large uncertainties such that a meaningful comparison can not be made at present.

\section{Strangeness fluctuations}
\begin{figure}[htb]
\centerline{%
  \includegraphics[width=12.5cm]{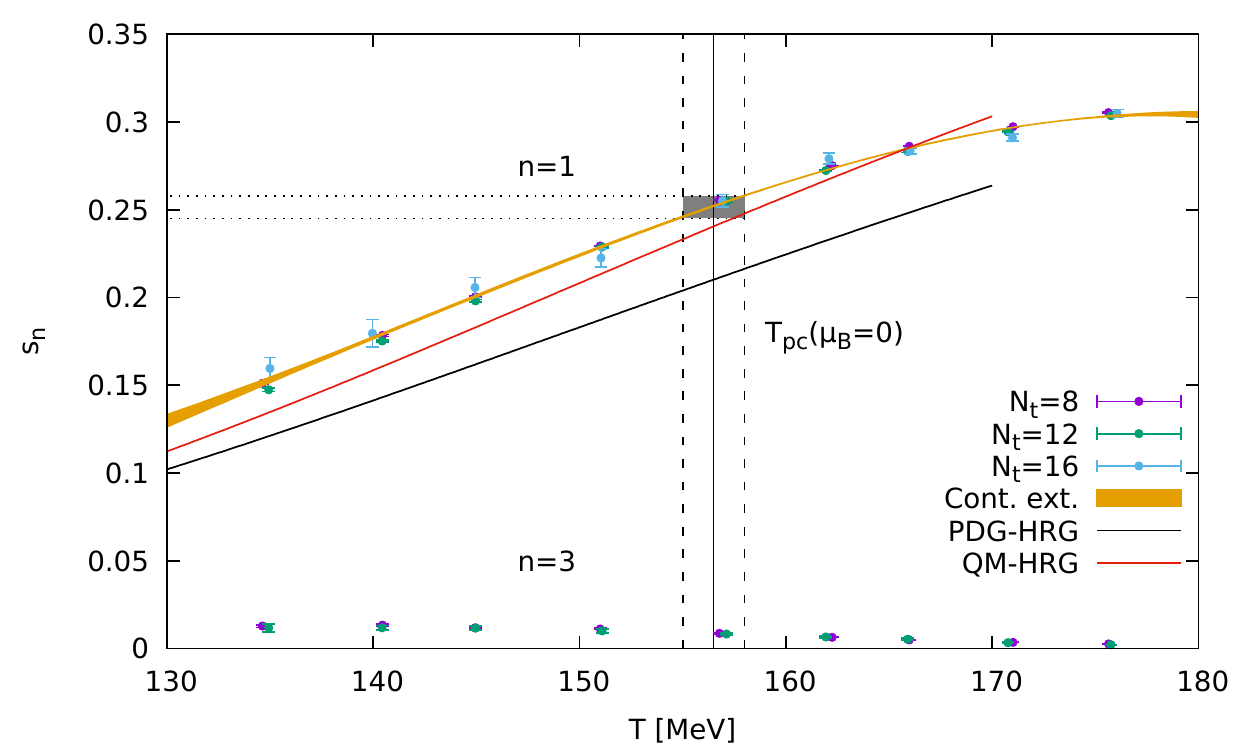}}
\caption{Temperature dependence of the expansion coefficients $s_1(T)$ and $s_3(T)$ for $N_t=8,12$ and $16$ lattices and PDG-HRG and QM-HRG curves for $s_1$.}
\label{fig:sfig}
\end{figure}

\begin{figure}[htb]
\centerline{%
  \includegraphics[width=12.5cm]{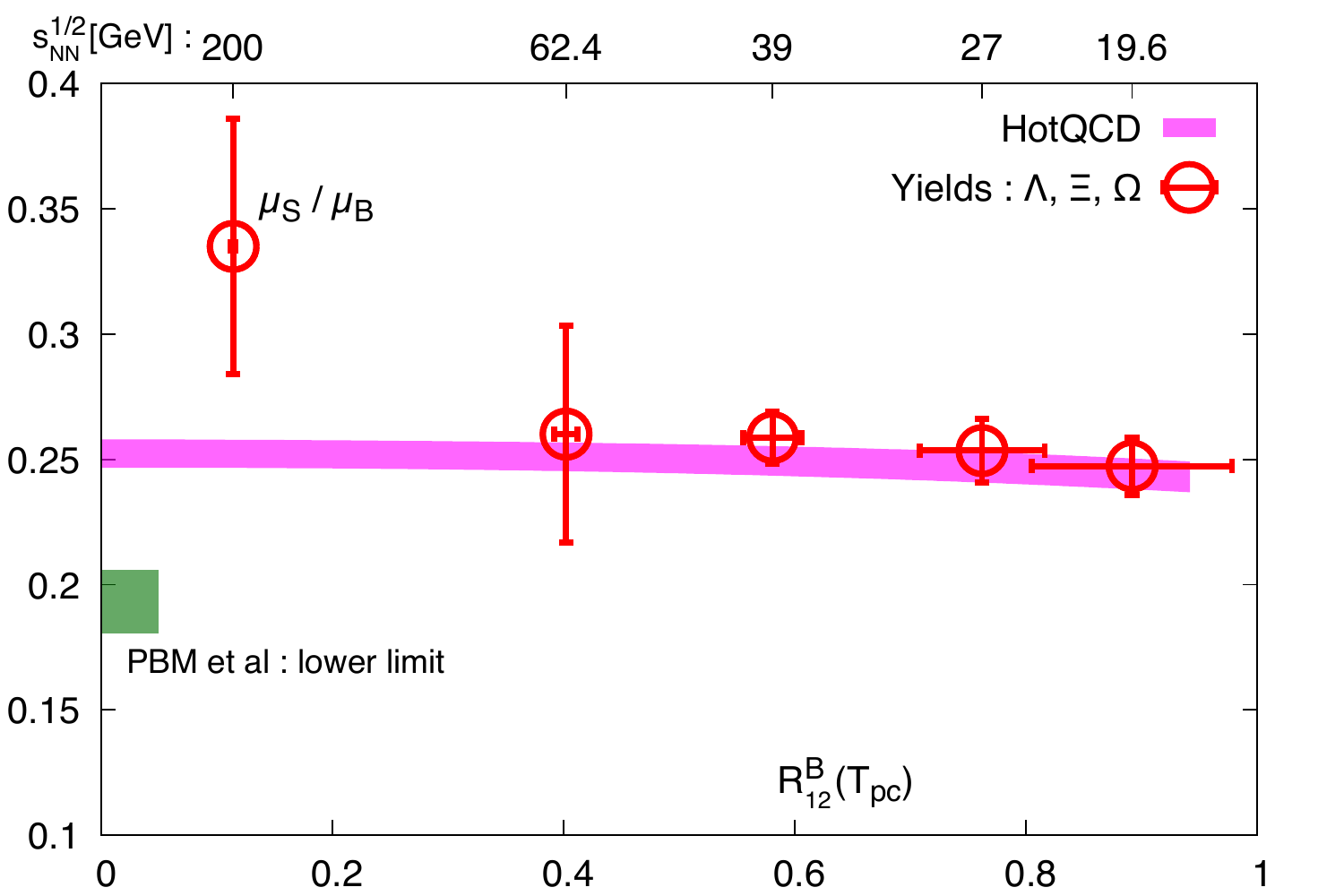}}
\caption{Comparison of $\mu_S/\mu_B$ from strange baryon yield fits and lattice QCD on the pseudo-critical line. The green box shows the lower limit on $-\chi^{BS}_{11}/\chi^{S}_{2}$ from \cite{pbm}.}
\label{fig:sfinal}
\end{figure}

While mean-to-variance, skewness and kurtosis ratios of strangeness fluctuations are also accessible via lattice QCD following the discussion presented in the previous section, we want to explore here a different application of strangeness observables that provides a helpful consistency check for estimated freeze-out parameters. Recall that the strangeness neutrality constraint fully determines the strangeness chemical potential $\mu_S(T,\mu_B)$ as seen in \eqref{eq:museries}. Rewriting this equation to obtain $\mu_S/\mu_B$ gives
\begin{align}
  \label{eq:strangemu}
  \frac{\mu_S}{\mu_B}=s_1(T)+s_3(T)\left(\frac{\mu_B}{T}\right)^2+\mathcal{O}\left(\left(\frac{\mu_B}{T}\right)^4\right).
\end{align}
This ratio is almost exclusively determined by $s_1(T)\approx -\frac{\chi^{BS}_{11}}{\chi^S_2}$. At the pseudo-critical temperature $T_{pc}(\mu_B=0)$, we find $s_1(T_{pc,0})=0.251(6)$. This is in agreement with a lower limit $-\frac{\chi^{BS}_{11}}{\chi^S_2} > 0.193 \pm 0.0127$ calculated in \cite{pbm} based on data from the ALICE experiment. Already the NLO coefficient $s_3$ is smaller than $s_1$ by an order of magnitude, as shown in Fig. \ref{fig:sfig}. $\mu_S/\mu_B$ is also sensitive to the strangeness content in hadron resonance gas models. The PDG-HRG, which contains the hadron states listed in the particle data booklet, shows a clear deviation from lattice QCD. This deviation shrinks if the QM-HRG, which contains additional, not yet observed hadronic states predicted by the quark model, is used instead. In a non-interacting hadron resonance gas, $\mu_S/\mu_B$ enters also in the ratio of anti-strange baryon to strange baryon yields $\bar{B}/B$ given by
\begin{align}
  \label{eq:hrg}
  \frac{\bar{B}}{B}(\sqrt{s})=\exp{\left(-\frac{\mu_B}{T}(2-2\vert S \vert \frac{\mu_S}{\mu_B})\right)}.
\end{align}
Assuming that HRG relations provide a good approximation for particle yields generated at the time of freeze out we may fit the experimentally measured yield ratios for different particle species in $|S|$ via (6). The ratio $\mu_S/\mu_B$ at $T=T_f$ can then be extracted from such a fit.
We performed this fit on the $\Lambda$, $\Xi$ and $\Omega$ yield data published by STAR in \cite{StarStrange1} and \cite{StarStrange2} and compare the result with $\mu_S/\mu_B$ from lattice QCD evaluated on the pseudo-critical line in Fig. \ref{fig:sfinal}. Apart from the data point at $\sqrt{s_{NN}}=200$ GeV, the strangeness to baryon chemical potential ratio at freeze-out, shown as red points, agrees very well with the lattice QCD result on the pseudo-critical line. 

\section{Summary}
We presented NNNLO calculations of the mean to variance ratio of electric charge fluctuations obtained from state of the art lattice QCD calculations and demonstrated how $R^Q_{12}$ can be used to extract freeze-out chemical potentials by comparing to measurements of this ratio by STAR. Furthermore, we estimated the skewness ratio $R^Q_{31}$ to NNLO in $\mu_B$ and found $R^Q_{31}(T_{pc}(\mu_B))=1.07(9)$ for $\mu_B<150$ MeV. This is in agreement with the skewness ratio measured by PHENIX for beam energies $27\;\mathrm{MeV} \leq \sqrt{s_{NN}} \leq 200 \;\mathrm{MeV}$. For the kurtosis ratio $R^Q_{42}$, we found $R^Q_{42}(T_{pc})=0.73(5)$ but large uncertainties in the measurement of the kurtosis ratio in experiments prevent comparisons at this time. Lastly, we used strangeness fluctuation observables to construct the ratio $\mu_S/\mu_B$ on the pseudo-critical line and compared it to $\mu_S/\mu_B$ at freeze-out determined by fitting strange hadron yields measured by STAR. Again, we found the results at freeze-out to be consistent with lattice QCD results on the pseudo-critical line.
\section{Acknowledgements}
This work was supported by the Deutsche Forschungsgemeinschaft (DFG, German Research Foundation) - project number 315477589 - TRR 211; the German Bundesministerium f\"ur Bildung und Forschung through Grant No. 05P2018 (ErUM-FSP T01) and the European Union H2020-MSCA-ITN-2018-813942 (EuroPLEx). It furthermore received support from the U.S. Department of Energy, Office of Science, Office of Nuclear Physics through (i) the Contract No. DE-SC0012704 and (ii) within the framework of the Beam Energy Scan Theory (BEST) Topical Collaboration, and (iii) the Office of Nuclear Physics and Office of Advanced Scientific Computing Research within the framework of Scientific Discovery through Advance Computing (SciDAC) award Computing the Properties of Matter with Leadership Computing Resources.
\bibliographystyle{unsrt}
\bibliography{chargefluc.bib}
\end{document}